\documentstyle[aps,graphicx,pre]{revtex}

\newcommand{\beas}{\begin{eqnarray*}}
\newcommand{\eeas}{\end{eqnarray*}}
\def\sign{\hbox{sign}\,}

\newcommand{\avg}[1]{\langle{#1}\rangle}
\newcommand{\Avg}[1]{\big\langle{#1}\big\rangle}
\newcommand{\ovl}[1]{\overline{#1}}
\newcommand{\BE}{\begin{eqnarray}}
\newcommand{\EE}{\end{eqnarray}}
\newcommand{\BEn}{\begin{eqnarray*}}
\newcommand{\EEn}{\end{eqnarray*}}
\newcommand{\barr}{\begin{array}}
\newcommand{\earr}{\end{array}}

\newcommand{\bit}{\begin{itemize}}
\newcommand{\eit}{\end{itemize}}
\newcommand{\bc}{\begin{center}}
\newcommand{\ec}{\end{center}}
\newcommand{\ben}{\begin{enumerate}}
\newcommand{\een}{\end{enumerate}}

\newcommand{\eps}{\epsilon}

\newcommand{\req}[1]{(\ref{#1})}
\newcommand{\be}{\begin{equation}}
\newcommand{\ee}{\end{equation}}
\newcommand{\bea}{\begin{eqnarray}}
\newcommand{\eea}{\end{eqnarray}}
\newcommand{\dd}{\textrm{d}}

\begin{document}
\twocolumn[\hsize\textwidth\columnwidth\hsize\csname
@twocolumnfalse\endcsname
\title{Criticality and finite size effects in a simple realistic model
of stock market}
\author{Damien Challet$^1$ and Matteo Marsili$^2$}
\address{$^1$ Theoretical Physics, Oxford University, 1 Keble Road, Oxford OX1 3NP, United Kingdom}
\address{$^2$ Istituto Nazionale per la Fisica della Materia (INFM),
Trieste-SISSA Unit,\\
Via Beirut 2-4, Trieste 34014, Italy}

\maketitle
\begin{abstract}
We discuss a simple model based on the Minority Game which reproduces
the main {\em stylized facts} of anomalous fluctuations in finance. We
present the analytic solution of the model in the thermodynamic limit
and show that stylized facts arise only close to a line of critical
points with non-trivial properties. By a simple argument, we show
that, in Minority Games, the emergence of critical fluctuations close
to the phase transition is governed by the interplay between the
signal to noise ratio and the system size. These results provide a
clear and consistent picture of financial markets as critical systems.
\end{abstract}
]

\narrowtext

Understanding the origin of the anomalous collective fluctuations
arising in stock markets poses novel and fascinating challenges in
statistical physics. Stock market prices are characterized by
anomalous collective fluctuations -- known as {\em stylized
facts}\cite{Daco} -- which are strongly reminiscent of critical
phenomena: Prices do not follow a simple random walk process, but
rather price increments are fat tailed distributed and their absolute
value exhibits long range auto-correlations, called volatility clustering.

The connection with critical phenomena is natural, because financial
markets are indeed complex systems of many interacting degrees of
freedom ---~the traders. By means of agent based modeling, it has been
realized \cite{market_models,SZ99,J99,CCMZ00,CMZ01} that stylized
facts are due to the way in which the trading activity of agents
interacting in a market ``dresses'' the fluctuations arising from
economic activity -- the so-called {\em fundamentals}.
Ref. \cite{CMZ01} has shown that very simple models based on the
Minority Game~\cite{CZ1} can reproduce a quite realistic and rich
behavior. Their simplicity makes an analytical approach to these
models possible, using tools of statistical physics.  Whether Minority
Games describe interacting traders is a matter of
debate~\cite{dollar,M01}. At any rate, the emergence of anomalous
fluctuations in these models, besides providing a scenario for the
behavior of real markets, poses questions in statistical physics which
deserve interest of their own.

In this Letter, we first introduce the simplest possible Grand
Canonical Minority Game (GCMG) which reproduces the main stylized
facts, i.e. fat tails and volatility clustering. Then we present the
analytic solution of this model in the relevant thermodynamic limit.
It shows that the behavior of GCMG, in this limit, exhibits Gaussian
fluctuation for all parameter values but on a line of critical points which
 marks a discontinuous phase transition.  For finite size systems, numerical simulations
reveal that stylized facts emerge close to the transition line, but
they abruptly disappear as the system size increases.
Remarkably, the vanishing of stylized facts when the system's size
increases also occurs in a variety of models of financial
markets~\cite{Stauffer}.  We present a theory of finite
size effects which is fully confirmed by numerical simulations. This
allows us to conclude that anomalous fluctuations are properties of
the critical point in GCMG. The phase transition is quite unique as it
mixes features which are typical of first order phase transitions --
as discontinuities and phase coexistence -- and of second order phase
transitions -- such as the divergence of correlation volumes and
finite size effects.


In the market described by the Minority Game~\cite{CZ1}, agents
$i=1,\cdots,N$ submit a bid $b_i(t)$ to the market in every period
$t=1,2,\ldots$. Agents whose bid has the opposite sign of the total
bid $A(t)=\sum_i b_i(t)$, win whereas the others lose. The bids of
agents depend on the value $\mu(t)$ of a public information variable,
which is drawn uniformly from the integers $1,\ldots, P$. In other
words, agents have {\em trading strategies} which prescribes to agent
$i$ a bid $a_i^\mu\pm 1$ for each information $\mu$. Each agent is
assigned one such strategy, randomly chosen from the set of $2^P$
possible strategies of this type. Agents are adaptive and may decide
to refrain from playing if their strategy is not good enough~\cite{SZ99,J99}.  More
precisely, the bids of agents take the form $b_i(t)=
\phi_i(t) a_i^{\mu(t)}$ where $\phi_i(t)=1$ or $0$ according to whether
agent $i$ trades or not. In order to assess the performance of their
strategy, agents assign scores $U_{i}(t)$ which they update by
\be
U_{i}(t+1)=U_{i}(t)-a_i^{\mu(t)}A(t)-\eps_i.
\label{Usi}
\ee
where
\be 
A(t)=\sum_{i=1}^N b_i(t)=\sum_{i=1}^N \phi_i(t) a_i^{\mu(t)}.
\label{At}
\ee 
So if $-a_i^{\mu(t)}A(t)$ is large enough, i.e., larger than $\eps_i$,
the score $U_i$ increases. The larger $U_i$, the more likely it is that
the agent trades ($\phi_i=1$). Here we suppose that~\cite{CamererHo}
\be 
{\rm Prob}\{\phi_i(t)=1\}
=\frac{1}{1+e^{\Gamma U_{i}(t)}}
\label{Logit}
\ee
where $\Gamma>0$ is a constant. 
A good strategy  prescribes bids $a_i^\mu$ which tend
to coincide with those $b(t)=-\sign A(t)$ of the minority of
agents.  The connection with markets goes along the lines of
Refs. \cite{J99,CCMZ00,CMZ01,M01}, which show that $A(t)$ is
proportional to the difference of price logarithms; here, we take
 $\log p(t+1)=\log p(t)+A(t)$.

The threshold $\eps_i$ in Eq. \req{Usi} models the incentives of
agents for trading in the market. Some investors may have incentives
to trade because they need the market for exchanging goods or
assets. This corresponds to $\eps_i<0$. On the contrary, speculators
who only trade for profiting of price fluctuations typically have
$\eps_i>0$. Of course there may be prudent investors with $\eps_i>0$
or risk-lover speculators with $\eps_i<0$ and a whole range of other
type of traders. Here we focus, for simplicity, on the case

\beas
\eps_i&=& \eps~~~~~\hbox{for $i\le N_s$}\\
\eps_i&=&-\infty~~~~~\hbox{for $N_s<i\le N$} 
\eeas 

\noindent
The $N_p=N-N_s$ agents who have $\eps_i=-\infty$ --- we call them {\em
producers} after Refs.~\cite{MMM,ZMEM} --- trade no matter what,
whereas the remaining $N_s$ --- the {\em speculators} --- trade only
if the cumulated performance of their active strategy increases more
rapidly than $\eps t$.


If the conditional time average $\avg{A|\mu}$ of $A(t)$ given
$\mu(t)=\mu$ is non-zero, then the knowledge of $\mu(t)$ allows a
statistical prediction of the sign of $A(t)$. A measure of
predictability is hence given by
\[
H_0=\ovl{\avg{A}^2}=\frac{1}{P}\sum_{\mu=1}^P \avg{A|\mu}^2
\]
where we introduced the notation $\ovl{(\ldots)}$ for averages over
$\mu$ ($\avg{\ldots}$ denotes averages on the stationary state). 
When $H_0=0$ the market is unpredictable or {\em
informationally efficient}. Volatility is instead defined as
$\sigma^2=\ovl{\avg{A^2}}$ and it measures market's fluctuations. A
further quantity of interest is the number $N_{\rm act}(t)=\sum_i
\avg{\phi_i(t)}$ of active speculators in the market. 

Exact results can be obtained in the thermodynamic limit, which is 
defined as the limit
$N_s,N_p,P\to\infty$, keeping constant the reduced number of
speculators and producers $n_s=N_s/P$, $n_p={N_p}/{P}$.
In this limit, both $\sigma^2$ and $H_0$ diverge with the system size,
since $A(t)\sim \sqrt{N}$. Hence we shall consider the rescaled
quantities $H_0/P$ or $\sigma^2/P$.  A detailed account of the
calculation will be given elsewhere
\cite{elsewhere}. Here we just discuss the main step and the results.
Following Ref. \cite{MC01}, we derive an Ito stochastic differential
equations for the strategy scores $y_i(\tau)= U_i(t)$ in the rescaled
continuous time $\tau= t/N$
\be
\frac{\dd y_i}{\dd \tau}=-\ovl{a_i\avg{A}_y}-\eps+\eta_i
\label{duidt}.
\label{langevin}
\ee 

\noindent
Here $\eta_i$ is a zero average Gaussian noise term with

\be
\avg{\eta_i(\tau)\eta_j(\tau')}=\frac{1}{N}
\ovl{a_i a_j \avg{A^2}_y}\delta(\tau-\tau').
\label{noise}
\ee

\noindent
In Eqs. (\ref{langevin},\ref{noise}) averages $\avg{\ldots}_y$ are
taken on the distribution of $\phi_i(t)$ in Eq. \req{Logit}, which
depends on $y_i(\tau)$ in a non-linear way: ${\rm Prob}\{\phi_i(t)=1\}
=1/[1+e^{\Gamma y_i(\tau)}]$. Hence Eq. \req{langevin}
is a quite complex system of non-linear equations with a noise
strength proportional to the time dependent volatility
$\ovl{\avg{A^2}_y}$. This feedback will be responsible for the
emergence of volatility build-ups.

Following Refs. \cite{CMZ,MC01} we find that the fraction
$\avg{\phi_i}$ of times that agent $i$ plays his active strategy in
the stationary state is the solution of the minimization of
the function 

\be 
H_\eps=\frac{1}{P}\sum_{\mu=1}^P
\left[\sum_{i=1}^N \avg{\phi_i} a_i^\mu+\sum_{i=N_s+1}^{N_s+N_p} a_i^\mu
\right]^2 +2\eps\sum_i \avg{\phi_i}
\ee
\noindent
with respect to $\avg{\phi_i}$. Note that for $\eps=0$ this function
reduces to the predictability $H_0$. For $\eps\neq 0$, the solution to
this problem, and hence the stationary state, is unique.  An exact
statistical mechanics description of the solution $\{\avg{\phi_i}\}$
can be carried out with the replica method, because the replica
symmetric ansatz is exact. Furthermore the solution to the
Fokker-Planck equation corresponding to Eq. \req{langevin} can be well
approximated by a factorized ansatz for $\eps\neq 0$. This means that
the off-diagonal correlations vanish [$\avg{(\phi_i-\avg{\phi_i})
(\phi_j-\avg{\phi_j})}=0$ for $i\neq j$] and, as a consequence, the
volatility turns out to be given by
$\sigma^2=\ovl{\avg{A^2}}=H_0+\sum_{i=1}^{N_s}
\avg{\phi_i}(1-\avg{\phi_i})$. The solution $\{\avg{\phi_i}\}$ of the
minimization of $H_\eps$ provides a complete description of the model
in the limit $N\to \infty$ for $\eps> 0$. In particular the
behavior is independent of $\Gamma$.

\begin{figure}
\includegraphics[width=7cm]{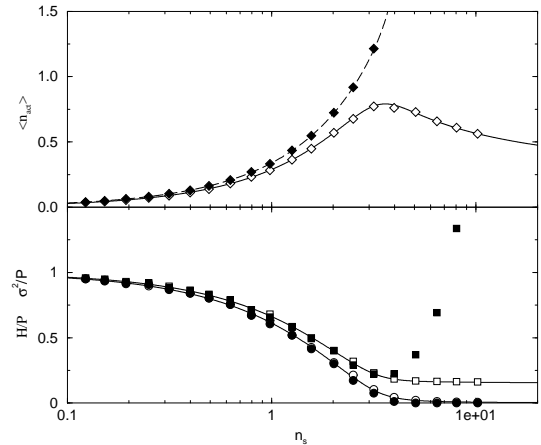}
\caption{Theory and numerical simulations:
$n_{\rm act}$ (top) and $\sigma^2/P$ and $H/P$ (bottom) as a function
of $n_s$ for $\eps=0.1$ (solid line) and $\eps=-0.01$ (dashed line).
Numerical results for $\eps=0.1$ (open symbols) and $\eps=-0.01$ (full
symbols) are averages over 200 runs, with
$N_sP=10000$ fixed and $\Gamma=\infty$.}
\label{figK}
\end{figure}

Fig. \ref{figK} shows that all these conclusions are perfectly
supported by numerical simulations: With a fixed number $n_p$ of
producers, as the number $n_s$ of speculators increases, the market
becomes more and more unpredictable, i.e. $H_0$ decreases. At the same
time also the volatility $\sigma^2$ decreases. In a market with few
speculators ($n_s<1$ in Fig. \ref{figK}), most of the fluctuations in
$A(t)$ are due to the random choice of $\mu(t)$ (i.e.  $\sigma^2\simeq
H_0$) and the number $n_{\rm act}$ of active speculators grows
approximately linearly with $n_s$.

When $n_s$ increases further, the market reaches a point where it is
barely predictable. Then, for $\eps>0$ the number of active traders
decreases and finally converges to
a constant. This means that the market becomes highly selective: Only
a negligible fraction of speculators trade ($\phi_i(t)=1$) whereas the
majority is inactive ($\phi_i(t)=0$). The volatility $\sigma^2$ also
remains constant in this limit.

For $\eps<0$ we see a markedly different behavior: The number of
active speculators continues growing with $n_s$ even if the market is
unpredictable $H_0\approx 0$. The volatility $\sigma^2$ has a minimum
and then it increases with $n_s$ in a way which depends on
$\Gamma$. In other words, $\eps=0$ for $n_s\ge n_s^\star(n_p)$
($=4.15\ldots$ for $n_p=1$) is the locus of a first order phase
transition across which $N_{\rm act}$ and $\sigma^2$ exhibit a
discontinuity. This same picture applies to a wider range of GCMG
models such as that of Ref. \cite{CMZ01}.

\begin{figure}
\includegraphics[width=7cm]{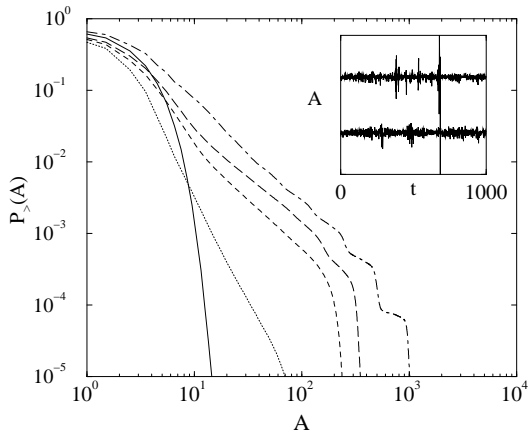}
\caption{Probability distribution of $A(t)>0$ for 
$n_s=10$ (continuous line), 20,50,100,200 (dash-dotted line) ($PN_s=16000$, $n_p=1$, $\eps=0.01$, $\Gamma=\infty$). 
Inset: time series of returns $A(t)$ showing volatility clustering for $n_s=20$ (lower curve), but not for $n_s=200$ (upper curve).}
\label{figpdf}
\end{figure}

Numerical
simulations reproduce anomalous fluctuations similar to those of real
financial markets close to the phase transition line.
 As shown in Fig. \ref{figpdf}, the distribution of
$A(t)$ is Gaussian for small enough $n_s$, and has fatter and fatter tails as $n_s$ increases; the same behavior is seen for decreasing $\eps$. 
In particular the distribution of $A(t)$ shows a power law
behavior $P(|A|>x)\sim x^{-\beta}$ with an exponent which we
estimated as $\beta\simeq 2.8, 1.4$ for $n_s=20, 200$ respectively
and $\eps=0.01$. 
Note that a
realistic value $\beta\approx 3$~\cite{RW} is obtained for $n_s=20$.

\begin{figure}
\includegraphics[width=7cm]{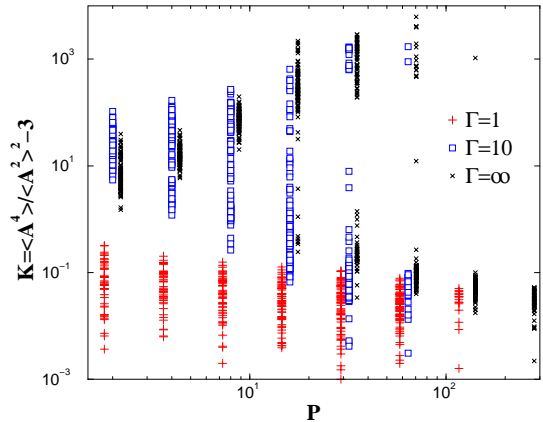}
\caption{Kurtosis of $A(t)$ in simulations with $\eps=0.01$, $n_s=70$,
$n_p=1$ and several different system sizes $P$ for $\Gamma=1,10$ and
$\infty$.}
\label{figkurt}
\end{figure}

This is inconsistent, at first sight, with the theoretical results
discussed previously for $N\to\infty$. Indeed, if the distribution of
$\phi_i$ factorizes, $A(t)$ is the sum of $N_s$ independent
contributions and it satisfies the Central Limit Theorem. This implies
that for $\eps\neq 0$ the variable $A(t)/\sqrt{N}$ converges in
distribution to a Gaussian variable with zero average and variance
$\sigma^2/N$ in the limit $N\to\infty$. There are no anomalous
fluctuations and no stylized facts. Fig. \ref{figkurt} indeed shows
that the anomalous fluctuations of Fig. \ref{figpdf} are finite size
effects which disappear abruptly as the system size increases (or if
$\Gamma$ is small).

\begin{figure}
\includegraphics[width=7cm]{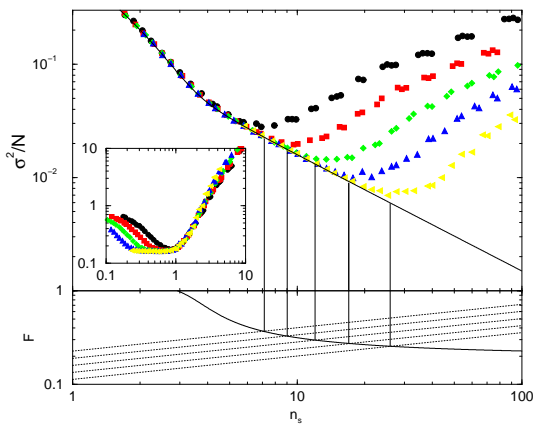}
\caption{Onset of the anomalous dynamics for different system sizes.
Top: $\sigma^2/N$ for different series of simulations with $L\equiv P
N_s$ constant: $PN_s=1000$ (circles), $2000$ (squares),
$4000$ (diamonds), $8000$ (up triangles) and $16000$ (left triangles).
 In all simulations $n_p=1$, $\eps=0.1$ and $\Gamma=\infty$.
Bottom: L.H.S. of Eq. \req{condvolclus} (full line) from 
the exact solution and $K/\sqrt{P}=K(n_s/L)^{1/4}$ (parallel dashed
lines) as a function of $n_s$ ($K\simeq 1.1132$ in this plot). 
The intersection defines $n_s^c(P)$. Inset: Collapse plot of 
$\sigma^2/N$ as a function of $n_s/n_s^c(P)$.}
\label{hdivsigma2}
\end{figure}

In order to understand these finite size effects, we note that
volatility clustering arises because the noise strength in
Eqs. (\ref{duidt},\ref{noise}) is proportional to the time dependent
volatility $\ovl{\avg{A^2}_{y}}$. The noise term is a source of
correlated fluctuations because $\ovl{a_ia_j\avg{A^2}_{y}}/N\sim
1/\sqrt{N}$ is small but non zero, for $i\neq j$.  It is reasonable to
assume that the dynamics will sustain collective correlated
fluctuations in the $y_i$ only if the correlated noise is larger than
the signal $-\ovl{a_i\Avg{A}_y}-\eps$ which agents receive form the
deterministic part of Eq. (\ref{duidt}). Time dependent volatility
fluctuations would be dissipated by the deterministic dynamics
otherwise. A quantitative translation of this insight goes as follows:
The noise correlation term is of order
$\ovl{a_ia_j\avg{A^2}_{y}}/N\sim\sigma^2/P^{3/2}$ for $i\neq j$. This
should be compared to the square of the deterministic term of
Eq. (\ref{duidt}) $[\ovl{a_i\avg{A}_{y}}+\eps]^2\sim
\left[\sqrt{H_0/P}+\eps\right]^2$. Rearranging terms, we find that
volatility clustering sets in when \be
\frac{H_0}{\sigma^2}+2\eps\sqrt\frac{H_0}{P}\frac{P}{\sigma^2}+
\eps^2\,\frac{P}{\sigma^2}\simeq\frac{K}{\sqrt{P}}
\label{condvolclus}
\ee
where $K$ is a constant. This prediction is remarkably well confirmed
by Fig. \ref{hdivsigma2}: In the lower panel we plot the two sides of
Eq. \req{condvolclus} as a function of $n_s$, for different system
sizes. The upper panel shows that the volatility $\sigma^2/N$ starts
deviating from the analytic result exactly at the crossing point
$n_s^c(P)$ where Eq. \req{condvolclus} holds true. Furthermore the
inset shows that the region $n_s>n_s^c(P)$ is described by a novel
type of scaling limit. Indeed the curves of Fig. \ref{hdivsigma2}
collapse one on top of the other when plotted against $n_s/n_s^c(P)$.

The non-linearity of the response of agents is crucial for the onset of
volatility time dependence. If $\Gamma$ is small the response becomes
smooth and anomalous fluctuations disappear (see
Fig. \ref{figkurt}). This picture is not affected by the
introduction of a finite memory in the learning process of agents, as
e.g. in Ref. \cite{MMRZ}. In particular the exponents of
Fig. \ref{figpdf} do not depend on the memory.

The fact that, in finite systems, stylized facts arise only close to
the phase transition is reminiscent of finite size scaling in the
theory of critical phenomena: In $d$-dimensional Ising model, for example, at
temperature $T=T_c+\varepsilon$ critical fluctuations (e.g. in the
magnetization) occur as long as the system size $N$ is smaller than
the correlation volume $\sim\varepsilon^{-d\nu}$. But for
$N\gg\varepsilon^{-d\nu}$ the system shows the normal fluctuations
of a paramagnet. 

Eq. \req{condvolclus} and $H_0/P\sim \eps^2$ imply that the same
occurs in the GCMG with $d\nu=4$. In other words, the critical window
shrinks as $N^{-1/4}$ when $N\to\infty$.  However, because of the long
range nature of the interaction, anomalous fluctuations either concern
the whole system or do not affect it at all, as clearly shown in
Fig. \ref{figkurt}. In the critical region the Gaussian phase coexists
probabilistically with a phase characterized by anomalous
fluctuations. This and the discontinuous nature of the transition at
$\eps=0$, are usually typical of first order phase transitions.

The picture of a phase transition controlled by the signal to noise
ratio appears to be universal for Minority Games. Finite size effects
close to the phase transition of the standard MG 
\cite{CZ1,elsewhere} are indeed explained by the same generic argument:
When the signal to noise ratio $H_0/\sigma^2$ is of order $1/\sqrt{P}$
self-sustained collective fluctuations arise.

Volatility clustering in real markets is known to be due to wild
fluctuations in the volume of trades \cite{RW}. Volume is the number
of active traders $N_{\rm act}+N_p$ in the GCMG.  Hence wild volume
fluctuations require correlated collective fluctuations in the
behavior of agents which only arise close to criticality. This
suggests that real markets operate close to a phase transition.
Numerical simulations suggest that exponents vary continuously on the
line of critical points. This raises the question of why real markets
self-organize close to the critical surface with $\alpha\approx 3$.

We conclude that the GCMG exhibits a quite peculiar type of phase
transition which mixes properties of continuous and discontinuous
transitions. Finite size effects clearly relate the occurrence of
stylized facts to the analytic nature of the phase transition. The
extension of renormalization group approaches to this system promises
to be a quite interesting challenge.


\begin{thebibliography}{99}
\bibitem{Daco} M. M. Dacorogna, {\em et al}.
{\em An Introduction to High-Frequency
Finance}, Chap. V, Academic Press, London (2001)
\bibitem{market_models} See e.g. B. LeBaron, W. B. Arthur,
R. G. Palmer, J. of Econ. Dyn. \& Control, {\bf 23}, (1999);
G. Caldarelli, M. Marsili and Y.-C. Zhang, Europhys. Lett. {\bf 40},
479 (1997); T. Lux and M. Marchesi, Nature {\bf 397}, 498 - 500
(1999); R. Cont and J.-P. Bouchaud, Macroecon. Dyn. {\bf 4}, 170
(2000); H. Levy, M. Levy, S. Solomon, {\em From Investor Behavior to
Market Phenomena}, Academic Press, London (2000) ; J. D. Farmer and
S. Joshi, Santa Fe Institute working paper 99-10-071.
\bibitem{SZ99} F. Slanina and Y.-C. Zhang, Physica A {\bf 272}, 257 (1999).
\bibitem{J99} P. Jefferies, M.L. Hart, P.M. Hui, N.F. Johnson,
Int. J. Th and Appl. Fin. {\bf 3}-3 (2000).
\bibitem{CCMZ00} D. Challet, A. Chessa, M. Marsili, Y.-C. Zhang,
Quant. Fin. {\bf1}, 168 (2001).
\bibitem{CMZ01} D. Challet, M. Marsili and Y.-C. Zhang, 
Physica A {\bf 294}, 514 (2001).
\bibitem{CZ1} Challet D. and Zhang Y.-C.,
Physica A {\bf 246}, 407 (1997) e-print adap-org/9708006
\bibitem{dollar} J. V. Andersen and D. Sornette, e-print  cond-mat/0205423 (2002).
\bibitem{GiardinaBouchaud} I. Giardina, J.-Ph. Bouchaud, cond-mat/0206222 (2002)
\bibitem{M01} M. Marsili, Physica A, {\bf 299}, 93 (2001).
\bibitem{Stauffer} E. Egenter, T. Lux, D. Stauffer, Physica A {\bf
268}, 250 (1999)
\bibitem{CamererHo} C. Camerer and T.-H. Ho, Econometrica {\bf 67},
827 (1999); A. Cavagna {\em et al}, Phys. Rev. Lett. {\bf 83}, 4429
(1999).
\bibitem{MMM}  Challet D., M. Marsili and Y.-C. Zhang, Physica A {\bf 
276}, 284 (2000).
\bibitem{ZMEM} Y.-C. Zhang, Physica A {\bf 269}, 30 (1999).
\bibitem{elsewhere} D. Challet and M. Marsili in preparation.
\bibitem{MC01} M. Marsili, D. Challet, Phys. Rev. E {\bf 64}, 056138
(2001).
\bibitem{CMZ} D. Challet, M. Marsili and R. Zecchina,
Phys. Rev. Lett. {\bf 84}, 1824 (2000).
\bibitem{MMRZ} M.  Marsili {\em et al}, Phys. Rev. Lett.  {\bf 87},
208701 (2001).
\bibitem{RW} V. Plerou, {\em et al}, Phys. Rev. E {\bf 62}, R3023
(2000).

\end{thebibliography}
\end{document}